# 2.856-GHz MODULATION OF CONVENTIONAL TRIODE ELECTRON GUN

S. J. Park, J. S. Oh, J. S. Bak, M. H. Cho, W. Namkung
Pohang Accelerator Laboratory, POSTECH, Pohang 790-784 Korea

*Abstract*

For the generation of picosecond (< 100 ps) electron beam pulses, we studied the RF modulation of a conventional triode electron gun. The feasibility study for this scheme has been experimentally investigated by modulating a triode gun of the Y-824 cathode-grid (KG) structure provided by the CPI Eimac, with 2.856-GHz pulsed RF's generated by a solid-state amplifier (SSA). In this paper, we present the methods and results of this investigation.

## 1 INTRODUCTION

Triode electron guns are used as the sources for linear colliders and injectors for storage rings. The essential component of a triode gun is the cathode-grid (KG) structure whose inter-electrode spacing is very short (~0.17 mm). The short inter-electrode spacing combined with the use of a high-voltage (HV) nanosecond pulser has made it possible to generate a single or multiple nanosecond electron beam pulses from the gun [1, 2]. The schematic diagram of a conventional triode gun is shown in Fig. 1.

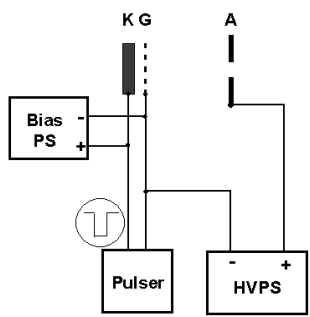

Fig. 1. Schematic diagram for conventional triode gun.

Generation of picosecond pulses from a conventional triode gun has not been practical because of a larger capacitance (~25 pF) of the KG structure and the unavailability of picosecond HV pulsers. In order to overcome these difficulties, we suggest the RF modulation of the KG structure. The advantage of this scheme becomes apparent when it is used as an electron source for an RF linac. If a triode gun is modulated at the same RF frequency as the accelerating structure, the bunching system including the pre-buncher and the buncher can be removed. This will greatly simplify the beginning part of a linac. Furthermore, the long-pulse (over several tens of microseconds) operation is possible, and it is beneficial for FEL experiments and industrial accelerators.

R. J. Becker and co-workers have modulated their triode gun by 1 GHz RF to obtain high-frequency long-pulse trains for the FELIX [3]. In order to analyze the RF modulation characteristics, they represented their KG structure as a simple resistance. This may become inaccurate as the RF frequency increases. The impedance of the KG structure becomes a combination of resistance, inductance, and capacitance. In order to evaluate the required RF power at the frequency of 2.856 GHz, a simple semi-analytic calculation of the RF modulation has been performed [4]. We sliced the electron beam starting from the cathode into a number of disks, and we solved the equation of motion with different initial RF phases. From this calculation, it was found that the RF power of about 100 W was required to be coupled to the KG structure for the peak beam current of 1 A at the modulation frequency of 2.856 GHz.

## 2 EXPERIMENTS

### 2.1 Zero-Beam-Voltage Test

In order to study the feasibility of the scheme, an experimental set-up has been prepared as shown in Fig. 2. In this set-up, no HV is applied between the KG structure and the anode. We tried to obtain the envelope of the modulated pulse train by measuring the current flowing out of the cathode.

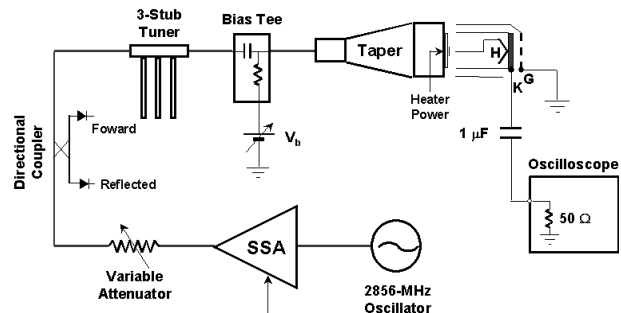

Fig. 2. Schematic diagram of experimental set-up (zero-beam-voltage test) for 2.856-GHz modulation of a conventional triode gun.

A 2856-MHz sinusoidal signal is fed into the SSA and amplified in a pulsed form. The pulsed RF power is then attenuated to a required level, and it is measured by a crystal detector attached to directional couplers. The impedance is matched by a coaxial tuner with 3-stubs. A special taper was fabricated to connect the Y-824 socket structure to the standard N-type connector. For feeding of heater current to the innermost heater socket, a cable should run across the inner and the outer conductors of the taper section. This point is located as close as to the Y-824 at which the voltage standing wave minimum is located. A Bias voltage between the cathode and the grid is provided through a tee. The insertion loss of the transmission line excluding the taper and 3-stub tuner was about 2.4 dB. The cathode current $I_K$, flowing from the cathode, was measured by an oscilloscope with 50-ohm input. A 1-µF capacitor is inserted to block out the DC bias. The $I_K$ was measured and plotted with varying heater voltage ($V_h$), bias voltage ($V_b$), and peak RF power ($P_{rf}$). The results are summarized in Fig. 3.

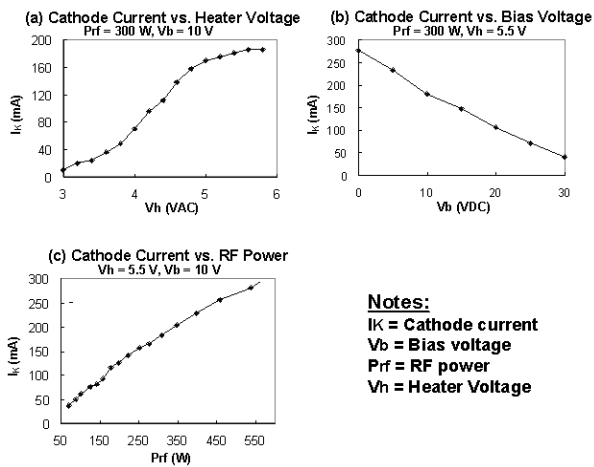

Fig. 3. $I_K$ measurements with varying $V_h$, $V_b$, and $P_{rf}$

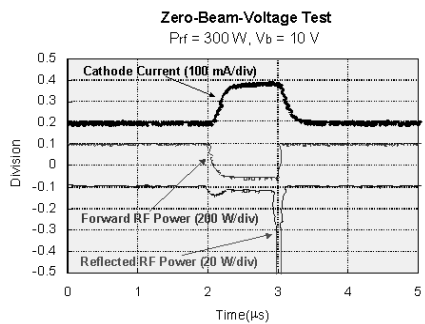

Fig. 4. Typical cathode current and RF powers waveforms

Figure 3(a) shows that the cathode current, $I_K$ saturates by the space-charge effect. Fig. 3(b) and 3(c) show the dependencies of the $I_K$ on the $V_b$ and the $P_{rf}$. The cathode current and RF power waveforms are shown in Fig. 4.

## 2.2 High-Voltage Test

As the next step, a HV was applied between the grid of the KG sructure and the anode. A HV deck was installed, and all the instruments and components were place on it. The schematic diagram of the set-up is shown in Fig. 5.

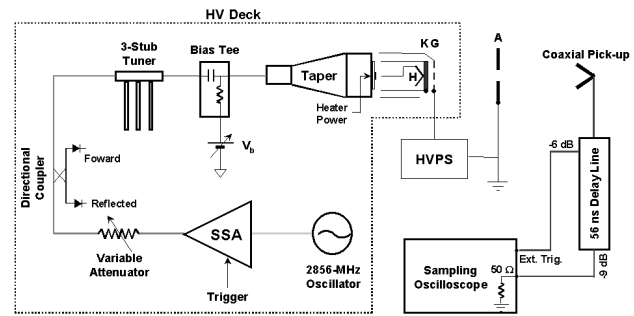

Fig. 5. Experimental set-up for high-voltage test of 2.856-GHz modulation of conventional triode gun.

The waveforms of electron beam pulses were measured by the use of a coaxial pick-up. A N-type coaxial feed-through was mounted on a 2.75″ CF flange. An Aluminium button was attached to the center conductor of the coaxial feedthrough as shown in Fig. 6.

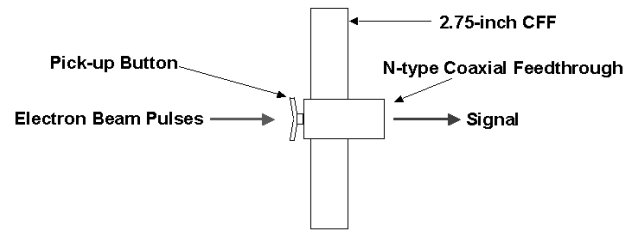

Fig. 6. Coaxial pick-up for measuring pulse waveforms of electron beams

The pick-up is meant only to measure the time structure of the electron beam pulses. The absolute measurement of the total charge is not our purpose. The capacitance of the button as measured by the Time Domain Reflecto-metry (TDR) technique was about 4 pF. Electrical pulse signals from the coaxial pick-up were measured using a 20-GHz sampling oscilloscope (Tektronix 11801B with SD-24 sampling head). A 56-ns delay line (Model 5056, Picosecond Pulse Labs.) was used to delay the pulse signals before entering the sampling oscilloscope. The delay line also provides trigger signals at −6 dB of the pulse signals. The bandwidth of all cables and connectors were about 20 GHz.

In Fig. 7, we show the measured electron beam pulse waveforms together with the input RF waveforms for comparison. Experimental conditions at the time of the measurement are as follows.

1. Acceleration HV = 50 kV,
2. Peak Input RF Power = 500 W,
3. RF Pulsewidth = 0.9 µs,
4. RF Pulse Repetition Frequency = 30 Hz,
5. Bias (Cathode-Grid) Voltage = 0 VDC.

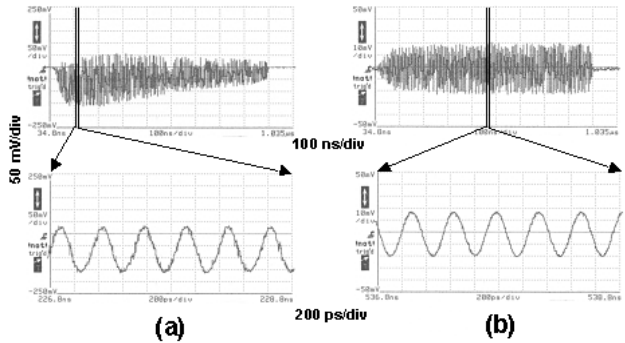

Fig. 7. (a) 2.856-GHz modulated pulsed-waveform for electron beams and (b) input RF waveforms. Top and bottom waveforms are macro and micro pulses, respectively.

The period of the measured waveforms was 350 ps corresponding to the RF frequency of 2.856 GHz. The shape of the waveform was almost sinusoidal with a negative offset. This is due to the limited response speed of the coaxial pick-up. Since the capacitance of the pick-up was 4 pF, the RC time is 50 ohms x 4 pF = 200 ps. This is too long to measure 100-ps pulses. The measured peak pulse-current was about 10 mA. This is not the total current because the pick-up button intercepts only a part of the total charge. Based on our experiences of nanosecond pulse measurements with the same coaxial pick-up, a rough estimation of the true pulse current is obtained by multiplying a factor of 10. Thus, the true peak pulse current could be roughly estimated to be 100 mA.

## 3 SUMMARY AND FURTHER WORK

The feasibility of the 2.856-GHz modulation of a conventional triode electron gun has been investigated experimentally. We summarize the results as follows.

1. Test with zero beam voltage showed that the cathode current in the KG structure under the 2.856-GHz RF modulation showed the space-charge saturation behaviour. Furthermore, the dependencies of the cathode current on the bias voltage and the RF power were as expected.
2. We measured 2.856-GHz modulated electron beam pulses. The period of the measured pulse waveform was 350 ps corresponding to the RF modulation frequency of 2.856 GHz. The peak pulse current was roughly 100 mA.
3. The feasibility of the 2.856-GHz modulation of a conventional triode gun has been demonstrated.

With these achievements, further work should include the followings.

1. Improvement of the response time of the coaxial pick-up.
2. Measurement of pulse waveform with varying operation parameters.
3. Increase of peak beam current up to 1 A.

In order to improve the response time of the coaxial pick-up, we are going to reduce the capacitance of the beam-collecting button below 2 pF. With this, it may be possible to measure 100-ps or less electron beam pulses. In order to increase the beam current, one may adopt a coaxial resonant cavity loaded by the KG structure. With this, the RF losses in the transmission lines could be minimized, and the efficient coupling of the RF power to the KG structure would be achieved.